\documentstyle[sprocl]{article}
\input{psfig}
\def\Journal#1#2#3#4{{#1} {\bf #2}, #3 (#4)}

\def\PLB{{\em Phys. Lett.}  B}

\def\PRD{{\em Phys. Rev.} D}
\def\PR{\em Phys. Rev.}
\def\ZPC{{\em Z. Phys.} C}
\def\ZP{\em Z. Phys.}
\def\be{\begin{equation}}
\def\ee{\end{equation}}
\def\bea{\begin{eqnarray}}
\def\eea{\end{eqnarray}}
\begin{document}
\title{PHOTOPRODUCTION OF JETS AT NLO$\;$\footnote{Presented at the 1996 
Meeting of the American Physical Society, Division of Particles and Fields 
(DPF 96), Minneapolis, Minnesota, 10-15 Aug 1996.}}
\author{BRIAN W. HARRIS and JOSEPH F. OWENS}
\address{Physics Department, Florida State University, \\ Tallahassee,
FL 32306-3016, USA}
\maketitle\abstracts{A new next-to-leading order Monte Carlo program for 
the calculation of fully differential jet cross sections in photoproduction 
is described.  The contributions from both resolved and direct components 
are included.  A comparison between the theoretical predictions and ZEUS 
data is presented.}

\section{Introduction}
The production of jets with large transverse momenta by photon beams 
differs in a number of ways from the corresponding hadroproduction case.  
These differences can be used to further investigate the interactions of 
the underlying hadronic constituents as well as to study the interactions 
between photons and hadrons \cite{owens}.  It has become conventional to 
describe jet photoproduction in terms of two components.  One is referred 
to as the direct component wherein the photon participates directly in the 
hard scattering subprocess.  The other is called the resolved component 
and corresponds to a situation where the photon interacts as if it 
contained partons.

\section{Method}
The calculation was performed using the phase space slicing method as 
was done in a previous study which included only the direct component 
at next-to-leading order (NLO) \cite{boo}.  
Briefly, two cutoffs are used to isolate the regions containing 
soft and collinear singularities from the remainder 
of the three-body phase space.  The three-body squared matrix elements 
are integrated over these singular regions analytically.  
The soft singularities cancel upon addition of the 
virtual contributions and the remaining collinear singularities are 
factorized into the parton distributions.  The integrations over 
the singularity free portions of the three-body 
phase space are performed using Monte Carlo methods.  When all of 
the contributions are combined at the histogramming stage the various 
cutoff dependences cancel, since the cutoffs merely mark the 
boundary between the region where the integrations were performed 
using analytic or Monte Carlo methods.

\begin{figure}
\centerline{\hbox{\psfig{figure=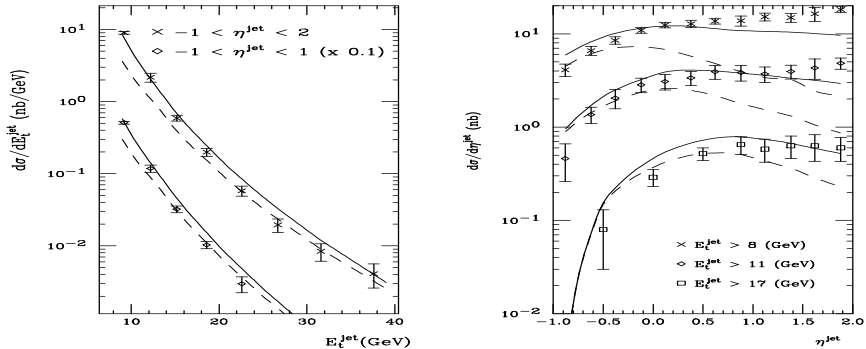,width=4.5in,height=1.8in}}}
\caption{Single inclusive jet cross section predictions compared to 
ZEUS \protect\cite{zeus} data.}
\end{figure}
\section{Results}
Results are presented for HERA where $27.5\,{\rm GeV}$ positrons are 
collided with $820\,{\rm GeV}$ protons.  Virtualities $Q^2$ 
of the photon are incorporated using the Weizs\"{a}cker-Williams 
approximation \cite{ww}.  We have used the two loop strong coupling 
with $n_f=5$, the proton distribution functions of CTEQ \cite{cteq}, 
and the photon distributions of GRV \cite{grv}, all in the 
${\overline {\rm MS}}$ scheme and all evaluated at 
$\mu_f=\mu_r=E_t^{max}/2$ where $E_t^{max}$ is the maximum $E_t^{jet}$ 
of the event.  The Snowmass jet convention \cite{snow} is used with $R=1$.

The separation between direct and resolved components is only well 
defined at leading order.  Following ZEUS, the fraction of the photon energy 
contributing to the production of the two highest transverse energy jets 
$x_{\gamma}^{OBS}$ is used to define these kinematic regions.  

Shown in Figure 1 are jet photoproduction cross sections for the 
kinematic range $0.2 < y < 0.85$ and $Q^2 < 4 {\rm GeV}^2$ as 
measured by ZEUS \cite{zeus} for various cuts on $E_t^{jet}$ and 
$\eta^{jet}$.  The solid lines are NLO QCD including 
both direct and resolved components.  The dashed line is direct only 
($x_{\gamma}^{OBS} \ge 0.75$).  It is clear from these results that 
both components are required by the data.

In Figure 2 $\cos\theta^*$ and $\bar{\eta}$ distributions from 
ZEUS \cite{zeus} are shown for dijet production.  Here 
$\bar{\eta}=(\eta^{jet_1}+\eta^{jet_2})/2$ and $\theta^*$ 
is the angle between the dijet axis and the beam axis in the dijet rest 
frame.  The steeper slope of the resolved component as compared to the 
direct component is readily apparent, as noted by ZEUS \cite{zeus}.  The 
higher order corrections to the $\bar{\eta}$ distribution for the direct 
component improve the agreement with the data relative to the leading order 
prediction shown as a dotted line.

\begin{figure}
\centerline{\hbox{\psfig{figure=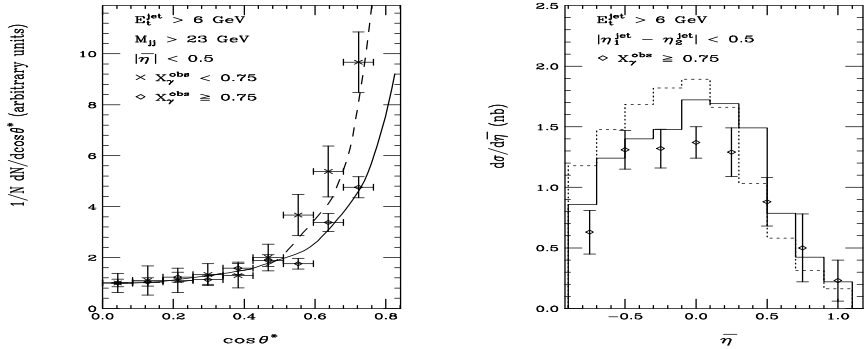,width=4.5in,height=1.8in}}}
\caption{Two jet cross section predictions compared to 
ZEUS \protect\cite{zeus} data.}
\end{figure}

\section{Conclusion}
Our results agree with the already available single inclusive results 
of Klasen {\it et al}. \cite{kl} and Aurenche {\it et al}. \cite{au}.  
Each of the examples given here has required placing cuts on several 
different variables.  The flexibility of the Monte Carlo approach has 
enabled us to calculate all of these simultaneously.  A thorough comparison 
with all available data is currently in progress.


\section*{References}
\begin{thebibliography}{99} 
\bibitem{owens}J.F. Owens, \Journal{\PRD}{21}{54}{1980}.
\bibitem{boo}H. Baer, J. Ohnemus and J.F. Owens, 
             \Journal{\PRD}{40}{2844}{1989}.
\bibitem{ww}C.F. Weizs\"{a}cker, \Journal{\ZP}{88}{612}{1934}; E.J. Williams, 
\Journal{\PR}{45}{729}{1934}.
\bibitem{cteq}H.L. Lai {\it et al}., \Journal{\PRD}{51}{4763}{1995}.
\bibitem{grv}M. Gl\"{u}ck, E. Reya and A. Vogt, \Journal{\PRD}{46}{1973}{1992}.
\bibitem{snow}J. Huth {\it et al}., Proc. of the 1990 DPF Summer Study on High 
Energy Physics, Snowmass, Colorado, edited by E.L. Berger (World Scientific, 
Singapore, 1992) p. 134.
\bibitem{zeus} ZEUS Collab., DESY 96-094; \Journal{\PLB}{342}{417}{1995}; 
              \Journal{\PLB}{348}{665}{1995}.
\bibitem{kl}M. Klasen, {\it hep-ph/9606453}, M. Klasen, G. Kramer and S.G. 
Salesch, \Journal{\ZPC}{68}{113}{1995} and references therein.
\bibitem{au}P. Aurenche {\it et al}., \Journal{\PLB}{338}{98}{1994} and 
references therein.
\end{thebibliography}
\end{document}